\documentclass[preprint,tightenlines,aps,floats,nofootinbib]{revtex4}
\usepackage{epsf}
\usepackage{graphicx}
\newcommand{\gsim}
{\;\raisebox{-.3em}{$\stackrel{\displaystyle >}{\sim}$}\;}
\newcommand{\lsim}
{\;\raisebox{-.3em}{$\stackrel{\displaystyle <}{\sim}$}\;}
\begin{document}

\title{Hiding the Higgs Boson With Multiple Scalars}
\author{Sally Dawson$^{a}$}
\email[]{dawson@bnl.gov}
\author{Wenbin Yan$^{b}$}
\email[]{wenbin.yan@stonybrook.edu}

\affiliation{
$^a$Department of Physics, Brookhaven National Laboratory,
Upton, NY 11973, USA\\
$^b$Department of Physics and Astronomy, SUNY Stony Brook, Stony Brook,
NY 11794, USA
\vspace*{.5in}}

\date{\today}

\begin{abstract}
We consider models
with multiple Higgs scalar gauge singlets and the resulting restrictions on
the parameters from precision electroweak measurements.
In these models, the scalar singlets mix with
the $SU(2)_L$ Higgs doublet, potentially
leading to reduced couplings of the scalars
to fermions and gauge bosons relative to the Standard Model
Higgs boson couplings.  Such models can
make the
Higgs sector difficult to explore at the LHC.
We emphasize the new physics resulting from the
addition of at least two scalar Higgs singlets.

\end{abstract}

\maketitle
\newpage

\section{Introduction}

One of the major goals of the Large Hadron Collider is to probe the
electroweak symmetry breaking sector.  The simplest implementation of
the symmetry breaking utilizes a single $SU(2)_L$ scalar Higgs doublet.  In
this minimal case, the couplings of the Higgs boson to fermions and
gauge bosons are fixed in terms of the particle masses
 and the phenomenology
has been extensively studied.  It is of interest, however,
to study extentions
of the  Higgs doublet model and to examine which
possibilities are allowed by current data and how
LHC Higgs phenomenology is affected.
  The most straightforward possibility  for enlarging
the Higgs sector is to
add some arbitrary number of scalar singlets which couple only to the
Higgs doublet.

The phenomenology of models with one scalar singlet in addition
to an $SU(2)_L$ doublet has been
examined by many
authors\cite{BahatTreidel:2006kx,O'Connell:2006wi,Barger:2007im,Profumo:2007wc,Bhattacharyya:2007pb,Barger:2008jx}.
The case with one
additional scalar is similar to that of models with a
radion\cite{Hewett:2002nk}.
For a supersymmetric model, the addition of a gauge singlet scalar
superfield leads to the NMSSM\cite{Barger:2006dh,Dermisek:2005gg,Dermisek:2005ar},
which solves the so-called ``$\mu$''
problem of the MSSM\cite{Ellis:1988er}.  Alternatively, scalar singlets have been advocated
as a signal for a hidden world   which interacts  only with the scalar
sector of the Standard
Model\cite{Bowen:2007ia,Schabinger:2005ei,Patt:2006fw,Strassler:2006im}.

In this paper, we consider non-supersymmetric models with
additional scalar gauge singlets.
In the case where there is a $Z_2$ symmetry in the scalar
sector, this class of theory
generically leads to a dark matter candidate, which is
the lightest scalar singlet.  Without   a $Z_2$ symmetry,
the scalar singlets can mix with the Standard Model Higgs doublet
and there is no dark matter candidate.
It is this alternative which we consider here.
The existence of multiple scalar singlets leads to changes in the scalar
interactions with gauge bosons and fermions.
Many  authors have considered the case where the lightest scalar has
a mass on the order of a few $GeV$
and attempted to
construct scenarios which
evade the LEP direct  Higgs production
bounds\cite{Chang:2008cw,O'Connell:2006wi}.  We consider an alternative
case where all the scalars are heavier than the LEP lower bound on the
Standard Model Higgs, $M_{H,SM}> 114 ~GeV$\cite{Barate:2003sz}.

The existence of scalars heavier than the LEP bound is restricted
by electroweak precision measurements\cite{Amsler:2008zzb,lepeww,Erler:2008ek}.
Since the dependence of the electroweak measurements on the scalar
masses is logarithmic, it is possible to make quite significant
changes in the scalar sector and still be consistent with
precision data\cite{Zhang:2006vt,Dugan:1991ck,Chen:2006pb}.  We
compare the predictions of models with multiple Higgs scalars with
the restrictions obtained
 from the $S,T$, and $U$ parameters\cite{Peskin:1991sw}.
We examine the cases with
one and two scalar singlets and
derive some general restrictions on the properties of models
with
extra scalar singlets.  In the Standard Model, precision
electroweak measurements restrict the Higgs mass to be less
than about $185~GeV$, $M_{H,SM} < 185~ GeV$\cite{lepeww}.
We consider the possibility of discovering
a Higgs-like boson with a mass significantly larger than
allowed in the Standard Model in a theory with multiple scalars.
As more and more singlets are added, the couplings of the
individual scalars
to the fermions and gauge bosons become weaker and weaker and
heavy scalars can potentially be compatible with precision
measurements.
We are motivated by the
analysis of Ref.\cite{BahatTreidel:2006kx} which attempted to hide
the Higgs signal at the LHC by introducing multiple Higgs
scalars. This reference concluded that a model with three scalars
with masses in the $120~GeV$ region
could evade discovery at the LHC\footnote{Ref. 
\cite{BahatTreidel:2006kx} found that a model with three scalars
with masses $m_0=118~GeV$, $m_1=124~GeV$,
and $m_2=130~GeV$ would elude detection at the LHC with $L=100~fb^{-1}$.}.

In Section \ref{themodel}, we summarize  the class of models
which we consider in this note, while Section
\ref{lims} contains our results for the $S$,
$T$, and $U$ parameters.  Our results for one and two  singlets
are contained in Section \ref{results},
along with  a discussion of the phenomenological implications of our
results for Higgs searches at the LHC.  Technical details
are summarized in two
appendices.   Section \ref{concs} contains  some conclusions.

\section{The Models}
\label{themodel}
We consider a class of models with $N$ scalar singlets, $S_i$, along with
an $SU(2)_L$ doublet, $H$,
\begin{equation}
H=\left(\begin{array}{c}
\Phi^+\\
{1\over\sqrt{2}}(h+v_H)
\end{array}\right), \,\,\quad  S_i=s_i+v_{s_i} \, .
\end{equation}
We assume that the scalar
potential is such that all scalars get a
vacuum expectation value (VEV),
\begin{eqnarray}
\langle H\rangle &=& {v_H\over \sqrt{2}}\nonumber \\
\langle S_i\rangle &=&v_{s_i} \, .
\end{eqnarray}
Since the singlets do not couple to the $SU(2)_L\times U(1)_Y$
gauge
bosons, they do not contribute to $M_W$ and $M_Z$ and hence $v_H$ must take
the Standard Model value, $v_H=246~GeV$. The VEVs are determined
from the scalar potential,
\begin{eqnarray}
V_{scalar}&=& \mu_H^2 \mid H\mid^2+\lambda_H(\mid H \mid^2 )^2
+\Sigma_i \mu_i \mid H\mid^2 S_i
+{1\over 2}\Sigma_{ij}\mu_{ij}^2 S_i S_j \mid H\mid^2\nonumber \\
&&+\Sigma_i M_i^3 S_i+\Sigma_{ij}M_{ij}^2 S_iS_j +
\Sigma_{ijk}\lambda_{ijk} S_i S_jS_k +
\Sigma_{ijkl}\lambda_{ijkl} S_i S_j S_k S_l\, .
\end{eqnarray}
Note that we make no assumptions about possible $Z_2$ symmetries in
the scalar sector and in general $h$ will mix with the $s_i$
scalars to form the mass eigenstates.

The $N+1$ scalar mass eigenstates are
defined to be $\phi_i,i=0...N$, with masses, $m_i$.  We assume
that $m_0$ is the lightest scalar. The mass eigenstates
are related to the
gauge eigenstates by an $(N+1)\times (N+1)$ unitary matrix $V$,
\begin{eqnarray}
\left(
\begin{array}{c}
\phi_{0}\\ \phi_{1}\\
. \\
.\\
.\\
\phi_N
\end{array}\right)
=V
\left(
\begin{array}{c}
h^{0}\\ s_{1}\\
. \\
.\\
.\\
s_N
\end{array}\right) \, .
\label{mixingdef}
\end{eqnarray}

Our results are expressed in terms of the elements of the
mixing matrix $V$, which can be calculated in any given model.
The couplings of the scalars
to the gauge bosons and fermions are\footnote{Note
that the Goldstone bosons have Standard Model couplings.}
\begin{eqnarray}
L&=&-\Sigma_{i=0,N}V_{0i}\phi_i\biggl\{{m_f\over v_H}\overline{f} f
+2 M_W^2 W_\mu^+W^{\mu -}+M_Z^2 Z_\mu Z^\mu\biggr\}
\nonumber \\
&&-{1\over 2v_H^2} \sum_{i,j=0,N}V_{0i}V_{0j} \phi_i \phi_j\biggl\{
2 M_W^2 W_\mu^+W^{\mu -} +M_Z^2 Z_\mu Z^\mu\biggr\}\, .
\end{eqnarray}
The production rates of the $\phi_i$ are suppressed
by $\mid V_{0i}\mid^2$ relative to the Standard Model Higgs
boson production rates. The branching ratios of the
lightest scalar, $\phi_0$, to Standard Model
particles are identical to the Standard Model branching ratios, while
the branching ratios for the heavier scalars depend on whether
the channels $\phi_i\rightarrow \phi_j\phi_k$ are
kinematically accesible\cite{O'Connell:2006wi}.

\section{Limits from Precision Electroweak Measurements}
\label{lims}
The limits on the parameters of
the scalar sector from precision electroweak measurements
can be studied
assuming that the dominant contributions
resulting from  the expanded scalar sector
 are to the gauge
boson 2-point functions\cite{Peskin:1991sw,Altarelli:1990zd},
$\Pi_{XY}^{\mu\nu}(p^2)=\Pi_{XY}(p^2)g^{\mu\nu}+B_{XY}(p^2)p^\mu p^\nu$,
with $XY=\gamma\gamma, \gamma Z, ZZ$ and $W^+W^-$.
We define the $S$,$T$ and $U$ functions
following the notation of Peskin and
Takeuchi\cite{Peskin:1991sw},
\begin{eqnarray}
\alpha S&=&
\biggl({4 s_\theta^2 c_\theta^2\over M_Z^2}\biggr)
\biggl\{ \Pi_{ZZ}(M_Z^2)- \Pi_{ZZ}(0)-
\Pi_{\gamma\gamma}(M_Z^2)
\nonumber \\ &&
-{c_\theta^2-s_\theta^2\over c_\theta s_\theta}\biggl(
\Pi_{\gamma Z}(M_Z^2)
-\Pi_{\gamma Z}(0)\biggr)\biggr\}\nonumber \\
\alpha  T &=&
 \biggl({ \Pi_{WW}(0)\over M_W^2}-{
\Pi_{ZZ}(0)\over M_Z^2}-{2 s_\theta
\over c_\theta }{\Pi_{\gamma Z}(0) \over M_Z^2}
\biggr)
\nonumber \\
\alpha U&=& 4 s_\theta^2\biggl\{
{ \Pi_{WW}(M_W^2)-\Pi_{WW}(0)\over M_W^2} -c_\theta^2
\biggl({ \Pi_{ZZ}(M_Z^2)-\Pi_{ZZ}(0)\over M_Z^2}\biggr)
\nonumber \\
&&-2 s_\theta c_\theta
\biggl(
{ \Pi_{\gamma Z}(M_Z^2)-\Pi_{\gamma Z}(0)
\over M_Z^2}\biggr)
-s_\theta^2
{ \Pi_{\gamma \gamma}(M_Z^2)\over M_Z^2}\biggr\} \, ,
\label{sdef}
\end{eqnarray}
where $s_\theta\equiv \sin\theta_W$ and $c_\theta\equiv \cos\theta_W$
and  any definition of $s_\theta$ can be used in Eq. \ref{sdef} since
the scheme dependence is higher order.

The scalar contributions
to $S, T$, and
$U$
from loops containing the $\phi_i$
 are gauge invariant\cite{Degrassi:1993kn} and can be  found from
Appendix 1 of Ref.
\cite{Chen:2008jg} or from Ref.
\cite{Hollik:1988ii}\footnote{The Standard Model contributions
to the gauge boson $2-$ point functions can
be found from Appendix 1 of Ref. \cite{Chen:2008jg}
by setting $\delta=\gamma=0$ and dropping the contributions
involving $K^0$ and $H^\pm$.
Our convention in this note for the sign of the 2-point functions is opposite
from that of Ref. \cite{Chen:2008jg}. }.
\begin{eqnarray}
%S&=&{1\over 12 \pi} \sum_{i=0,N}\biggl\{\log\biggl({m_i^2\over \mu^2}\biggr)
%+{(9m_i^2+M_Z^2)M_Z^4\over (m_i^2-M_Z^2)^3}\log
%\biggl({m_i^2\over M_Z^2}\biggr)
%-{5m_i^4+14m_i^2M_Z^2+41 M_Z^2\over 6 (m_i^2-M_Z^2)^2}
S_\phi&=&
{1\over \pi}\Sigma_i\mid V_{0i}\mid^2\biggl\{
B_0(0,m_i,M_Z)- B_0(M_Z,m_i,M_Z)
\nonumber \\ &&
+{1\over M_Z^2}\biggl[
B_{22}(M_Z,m_i,M_Z)-B_{22}(0,m_i,M_Z)\biggr]\biggr\}
\nonumber \\
T_\phi&=& {1\over 4\pi s_\theta^2}
\Sigma_i\mid V_{0i}\mid^2
\biggl\{- B_0(0,m_i,M_W)+{1\over c_\theta^2}B_0(0,m_i,M_Z)\nonumber \\
&&
+{1\over M_W^2}\biggl(B_{22}(0,m_i,M_W)-B_{22}(0,m_i,M_Z)
\biggr)\biggr\}\nonumber \\
(U+S)_\phi&=& {1\over \pi}
\Sigma_i\mid V_{0i}\mid^2\
\biggl\{
B_0(0,m_i,M_W)-B_0(M_W,m_i,M_W)
\nonumber \\ &&
+{1\over M_W^2}\biggl[-B_{22}(0,m_i,M_W)
+B_{22}(M_W,m_i,M_W)\biggr]\biggr\}\, .
\label{stuphi}
\end{eqnarray}
The definitions of the Passarino-Veltman $B$ functions are
given in
 Appendix A.  The contributions from the Goldstone bosons are
identical to the Standard Model case and hence are not included
in Eq. \ref{stuphi}\footnote{ Eq. \ref{stuphi} is in agreement with
the results
of Ref. \cite{Profumo:2007wc} when the Goldstone boson contributions
are included.}.

Using the results in Appendix A,
\begin{eqnarray}
S_\phi&=&
{1\over \pi}\Sigma_i\mid V_{0i}\mid^2\
\biggl\{
-{1\over 8}{m_i^2\over M_Z^2}
+{m_i^2 \over m_i^2-M_Z^2}
\biggl(1-{m_i^2\over 4 M_Z^2}
\biggr)
\ln\biggl({M_Z^2\over m_i^2}\biggr)
\nonumber \\
&&+F_1(M_Z^2,m_i,M_Z)-{m^2_i\over 2 M_Z^2}F_2(M_Z^2,m_i,M_Z)
+C_S\biggr\}
\nonumber \\
T_\phi&=& -{3\over 16\pi s_\theta^2}
\Sigma_i\mid V_{0i}\mid^2\
\biggl\{
{m_i^2\over m_i^2-M_W^2}\ln\biggl({M_W^2\over m_i^2}\biggr)
-
{m_i^2\over c_\theta^2
(m_i^2-M_Z^2)}\ln\biggl({M_Z^2\over m_i^2}\biggr)
+C_T\biggr\}
\nonumber \\
(U+S)_\phi&=&
{1\over \pi}\Sigma_i\mid V_{0i}\mid^2\
\biggl\{
-{1\over 8}{m_i^2\over M_W^2}
+{m_i^2 \over M_i^2-M_W^2}
\biggl(1-{m_i^2\over 4 M_W^2}
\biggr)
\ln\biggl({M_W^2\over m_i^2}\biggr)
\nonumber \\
&&+F_1(M_W^2,m_i,M_W)-{m^2_i\over 2 M_W^2}F_2(M_W^2,m_i,M_W)
+C_U\biggr\}\, ,
\end{eqnarray}
where the terms $C_S, C_T$ and $C_U$ represent contributions which are
independent of $m_i$.

In order to compare with fits to data,
 we must   subtract from Eq. \ref{stuphi} the Standard
Model  Higgs
boson
contribution evaluated at a reference Higgs mass, $M_{H,ref}$,
\begin{eqnarray}
S_{H,ref}&=&
{1\over \pi}
\biggl\{
-{1\over 8}{M_{H,ref}^2\over M_Z^2}
+{M_{H,ref}^2 \over M_{H,ref}^2-M_Z^2}
\biggl(1-{M_{H,ref}^2\over 4 M_Z^2}
\biggr)
\ln\biggl({M_Z^2\over M_{H,ref}^2}\biggr)
\nonumber \\
&&+F_1(M_Z^2,M_{H,ref},M_Z)-{M^2_{H,ref}\over 2
M_Z^2}F_2(M_Z^2,M_{H,ref},M_Z) +C_S\biggr\}
\nonumber \\
T_{H,ref}&=& -{3\over 16\pi s_\theta^2}
\biggl\{
{M_{H,ref}^2\over M_{H,ref}^2-M_W^2}
\ln\biggl({M_W^2\over M_{H,ref}^2}\biggr)
-
{M_{H,ref}^2\over c_\theta^2 (M_{H,ref}^2-M_Z^2)}
\ln\biggl({M_Z^2\over M_{H,ref}^2}\biggr)
+C_T\biggr\}
\nonumber \\
(U+S)_{H,ref}&=&
{1\over \pi}
\biggl\{
-{1\over 8}{M_{H,ref}^2\over M_W^2}
+{M_{H,ref}^2\over M_{H,ref}^2-M_W^2}
\biggl(1-{M_{H,ref}^2\over 4 M_W^2}
\biggr)
\ln\biggl({M_W^2\over M_{H,ref}^2}\biggr)
\nonumber \\
&&+F_1(M_W^2,M_{H,ref},M_W)-{M^2_{H,ref}\over 2
M_W^2}F_2(M_W^2,M_{H,ref},M_W) +C_U\biggr\}\, . \label{stusm}
\end{eqnarray}
Finally, we compare the quantities from Eqs.
\ref{stuphi} and \ref{stusm},
\begin{eqnarray}
\Delta S_\phi&=&S_\phi-S_{H,ref}
\nonumber \\
\Delta T_\phi&=&T_\phi-T_{H,ref}
\nonumber \\
\Delta U_\phi&=&U_\phi-U_{H,ref}\, ,
\label{stufits}
\end{eqnarray}
with a fit to experimental data in order to obtain limits on
the allowed masses and mixing angles. For $m_i, M_{H,ref} >> M_W, M_Z$,
we find the familiar forms\cite{Peskin:1991sw},
\begin{eqnarray}
\Delta S_\phi&=&{1\over 12 \pi}
\Sigma_i\mid V_{0i}\mid^2
\log\biggl({m_i^2\over M_{H,ref}^2}\biggr)
\nonumber \\
\Delta T_\phi&=& -{3\over 16 \pi c_\theta^2}
\Sigma_i\mid V_{0i}\mid^2\log\biggl({m_i^2\over M_{H,ref}^2}\biggr)
\nonumber \\
\Delta U_\phi&=& 0
\, .
\label{loglim}
\end{eqnarray}
For $m_i\sim M_W,M_Z$ the ${\cal O}({M_W^2\over m_i^2},{M_Z^2\over m_i^2})$
terms which are neglected in
Eq. \ref{loglim} are numerically
important\footnote{Ref.\cite{BahatTreidel:2006kx} retains only the
logarithmic contributions.}.
Our fitting proceedure includes the complete result and
is described in Appendix B.

\section{Results}
\label{results}
In this section, we consider models with one and two scalar
singlets in addition to the $SU(2)_L$ doublet,
and extract the regions of parameter space allowed
by precision electroweak measurements.  The goal is to draw some
general conclusions about the Higgs discovery potential in models with 
expanded
scalar sectors.

The dominant discovery channel for  much of the Higgs mass
range is $\phi_i\rightarrow Z Z^* \rightarrow$ 4 leptons.  The
production rates of the  $\phi_i$ are
reduced from the Standard Model rates by $\mid V_{0i}\mid^2$.
For $m_i \lsim 200~GeV$, the $\phi_i$ scalar decay
 width is less than or
comparable to the detector resolution\cite{Bayatian:2006zz,Aad:2009wy},
so we use the narrow width
approximation and neglect effects of the finite scalar widths.
For the lightest Higgs boson, $\phi_0$, the Higgs branching ratios
are identical to the Standard Model branching ratios.  For the
heavier Higgs bosons, the scalar branching ratios depend on
whether the $\phi_i\rightarrow \phi_j\phi_k$ channel is accessible
 for
some $\phi_j$ and $\phi_k$.  Whether or not this channel is open depends
on the scalar mass spectrum, along with the parameters of the
scalar potential.  We define $\zeta_{ijk}=1 (0) $ if the decay
 $\phi_i\rightarrow \phi_j\phi_k$ is (is not) allowed.   The signal
for $\phi_i$ production with the subsequent decay to Standard
Model particles is then suppressed from the Standard Model rate
by\cite{O'Connell:2006wi},
\begin{eqnarray}
X_i^2&=&\mid V_{0i}\mid^2{\mid V_{0i}\mid^2\Gamma_h^{SM}\over
\mid V_{0i}\mid^2\Gamma_h^{SM} +\Sigma_{jk}\zeta_{ijk}
\Gamma (\phi_i\rightarrow \phi_j\phi_k)}\, ,
\label{reduce}
\end{eqnarray}
where $\Gamma_h^{SM}$ is the total  width in the Standard Model
for a Higgs boson of mass $m_i$.
For $\zeta_{ijk}=0$, $X_i=V_{0i}$.  From Eq. \ref{reduce}, $X_i$ is
always less than one, so the addition of scalar singlets reduces the
significance of the usual Higgs discovery channels.

\begin{figure}[t]
\begin{center}
\includegraphics[scale=0.8]{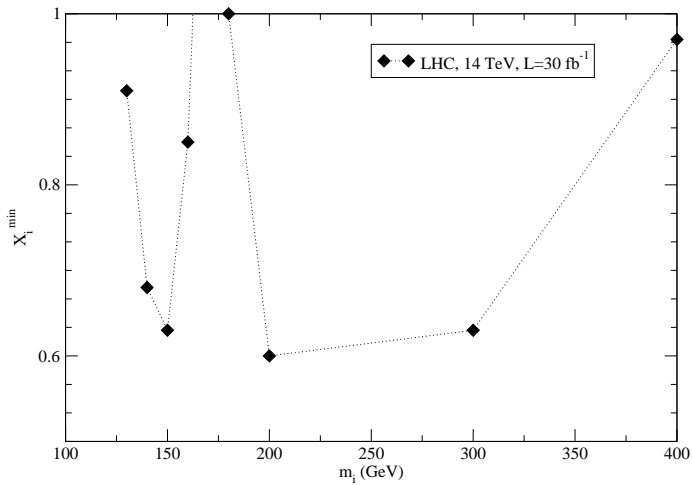}
\caption[]{Minimum value of $X_i$ for which a $5\sigma$
significance in the $\phi_i\rightarrow Z Z^* \rightarrow$ 4 lepton
channel is obtained
with the ATLAS detector
at the LHC with $\sqrt{s}=14~TeV$ and $\int L=30~fb^{-1}$\cite{Aad:2009wy}. }
\label{fg:atlim}
\end{center}
\end{figure}

  Fig.
\ref{fg:atlim} shows the minimum value of $X_i$,  $X_i^{min}$,
 for which a $5\sigma$
significance in the $\phi_i \rightarrow Z Z^* \rightarrow $ 4 lepton
channel can be found at the LHC with $\sqrt{s}=14~TeV$
and $L=30~fb^{-1}$.  This figure is  obtained by
rescaling recent ATLAS studies\cite{Aad:2009wy}.
As long as the $\phi_i\rightarrow \phi_j \phi_k$ channel is
closed for the heavier scalars,
then this limit can be trivially applied for all $\phi_i$ and 
$V_{0i}$\footnote{This also requires
that the mass differences between the scalars be greater
than the detector resolution.}.
For a model with one singlet and one $SU(2)_L$
doublet (and hence 2 physical Higgs bosons),
if both scalars have masses less than $m_{\phi_i}\sim 160~GeV$, then
since at least one of the scalars must have $V_{0i}>1/\sqrt{2}$,
at least one scalar can
be discovered through the $\phi_i\rightarrow Z Z^* \rightarrow$ 4 lepton
channel.  The situation changes when a second  singlet is added.
Now there are three physical scalars and it is possible for all scalars
to have masses less than $\sim 160~GeV$ and to have mixing angles $V_{0i}
\sim {1/\sqrt{3}}$.  In this case none of the scalars will be seen
(at least with  $L=30~fb^{-1}$) in the
$\phi_i\rightarrow Z Z^* \rightarrow$ 4 lepton
channel.  This is a generalization of the
result of Ref. \cite{BahatTreidel:2006kx} and can be
straight-forwardly applied to examples with more singlets.

In the region $165~GeV \lsim m_i \lsim 180~GeV$, the
$\phi_i\rightarrow Z Z^* \rightarrow$ 4 lepton channel does not
lead to a $5\sigma$ discovery with $30~fb^{-1}$.  In this mass
region, the most useful discovery channel is $\phi_i\rightarrow
W^+W^-\rightarrow e^\pm \nu \mu^\mp \nu $, which yields a $>5\sigma$ discovery
for $140~GeV \lsim m_i\lsim 185~GeV$ for $X_i\sim 1$\cite{Aad:2009wy}.
For $m_i \sim~160~GeV$, a $5\sigma$ discovery is possible with $X_i \gsim
0.7$.

\subsection{Fit with One Singlet}
\begin{figure}[t]
\begin{center}
\includegraphics[scale=0.75]{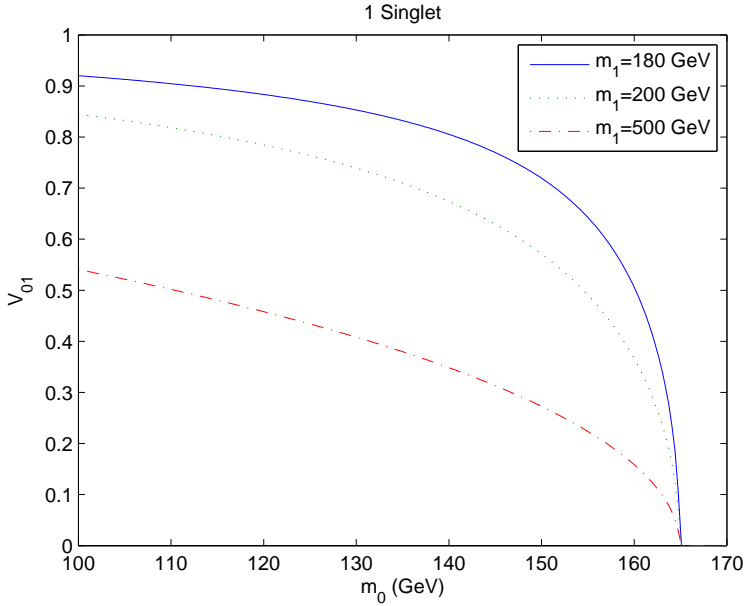}
\caption[]{Allowed region (at $95\%$ confidence level) in a model
with one additional singlet in addition to the usual $SU(2)_L$
doublet.  The lightest (heavier) scalar is $m_0$ ($m_1$) and the
mixing matrix is defined in Eq. \ref{mixingdef}. The region below
the curves is allowed by fits to $S,T$ and $U$.} \label{fg:singsa}
\end{center}
\end{figure}

\begin{figure}[t]
\begin{center}
\includegraphics[scale=0.8]{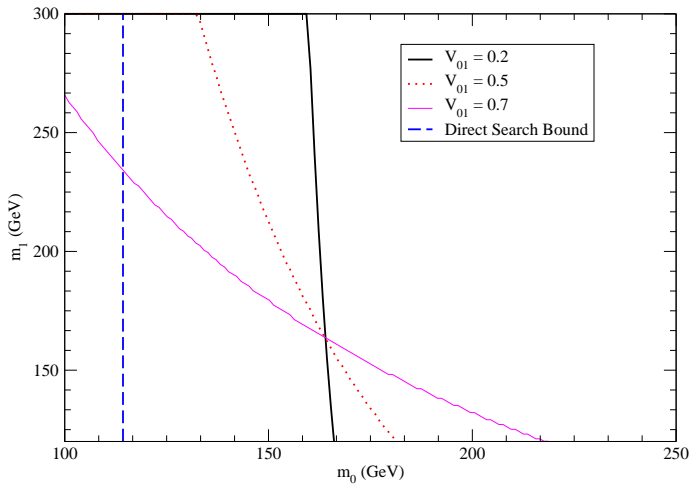}
\caption[]{Allowed region (at $95\%$ confidence level) in a model with
one additional singlet in addition to the usual $SU(2)_L$ doublet.  The
lightest (heavier) scalar is $m_0$ ($m_1$) and the mixing matrix is
defined in Eq. \ref{mixingdef}. The region below and
to the left of the curves labelled
with values of $V_{01}$  are allowed by
fits to $S,T$ and $U$.  The region to the right of the dashed line is allowed
by direct search limits from LEP2.}
\label{fg:singsb}
\end{center}
\end{figure}
Figs.~\ref{fg:singsa} and \ref{fg:singsb}
show results with one singlet scalar in addition to the $SU(2)_L$
scalar doublet.  The scalar sector is described by the masses of the
two scalars, $m_0$ and $m_1$, and one mixing angle which we take
to be $V_{01}$ ($V_{00}=\sqrt{1-V_{01}^2}$).
The fit to the experimental limits on $\Delta S$,
$\Delta T$ and $\Delta U$ is performed as described in Appendix B and
the maximum allowed value of $V_{01}$ for various
values of $m_0$ is shown in Fig.
\ref{fg:singsa}.   (For simplicity, we assume $\zeta_{ijk}=0$ for
all $i,j,k$.)
For $V_{01}\sim 0$, $\phi_0$ is
predominantly the neutral component of
the $SU(2)_L$ doublet with nearly Standard Model
couplings and the $95\%$~confidence
level limit
on the allowed value of $m_0$ is just
the $95~\%$ confidence level limit
of this fit in the Standard Model, $M_{H,SM}\lsim 166~GeV$. There is no limit
on $m_1$ in this case.

For moderate mixing, the heavier scalar, $\phi_1$,
can be quite  heavy.  For example, the lightest scalar could have
$m_0 \sim 140~GeV$ with a coupling $V_{00}\sim .7$ while the heavier scalar
could have a mass
$m_1\sim 200~GeV$ with a coupling $V_{01}\sim .7$.  In this case, comparison
with Fig. \ref{fg:atlim} shows that both scalars could be observed
in the $ZZ^*\rightarrow 4$ lepton channel with $30~fb^{-1}$.
If $\phi_1$ becomes too heavy (say $\phi_1\sim 500~GeV$), then its coupling
to Standard Model particles is
 restricted by the precision
electroweak measurements to be less than $V_{01}\lsim
0.5$ (for $m_0 \gsim 114~GeV$) and so $\phi_1$ cannot
be found in the $ZZ^*\rightarrow 4$~lepton mode with $10~fb^{-1}$.

Fig. \ref{fg:singsb} demonstrates that scalars which have masses in the
$200~GeV$ range can be compatible with the electroweak precision
measurements and have couplings large enough to be discovered at the LHC.
In much of the parameter space of this plot, both scalars will be
observed.

\subsection{Fit with Two Singlets}
In this subsection, we examine how the allowed masses of the scalars
are changed with the addition of two singlets in addition
to the Standard Model doublet. The scalar sector now has three
scalars with masses $m_0$, $m_1$ and $m_2$ and the mixing matrix
$V$ is a $3\times 3$ unitary matrix.  The phenomenology is quite
different from the case with one singlet.  As mentioned previously,
with two singlets it is possible to sufficiently suppress the couplings
$V_{0i}$ to all scalars such that none of them are observable
with $10~fb^{-1}$ at the LHC if they
all satisfy the Standard Model limit, $m_i\lsim 166~GeV$.

\begin{figure}[t]
\begin{center}
\includegraphics[scale=0.75]{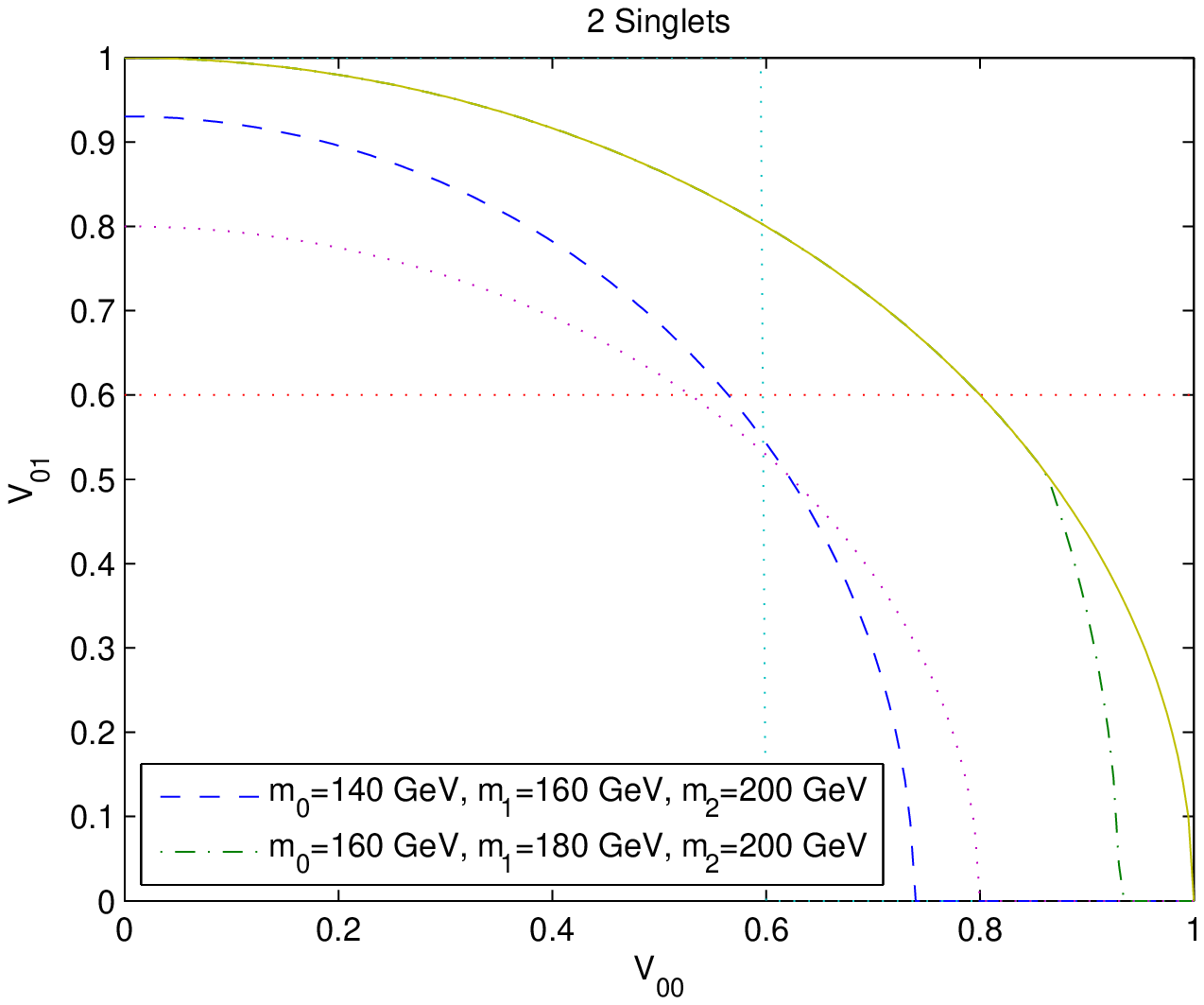}
\caption[]{Allowed region at $95\%$ confidence level for a model
with two singlets in addition to the $SU(2)_L$ scalar doublet.  The
allowed regions are above and to the right of the dashed and
dot-dashed curves. The solid curve is $\sum_i|V_{0i}|^2=1$. The
curved dotted line is $V_{02}=0.6$, while
the straight dotted lines are $V_{00}=0.6$ 
and $V_{01}=0.6$.} 
\label{sing2a}
\end{center}
\end{figure}

\begin{figure}[t]
\begin{center}
\includegraphics[scale=0.75]{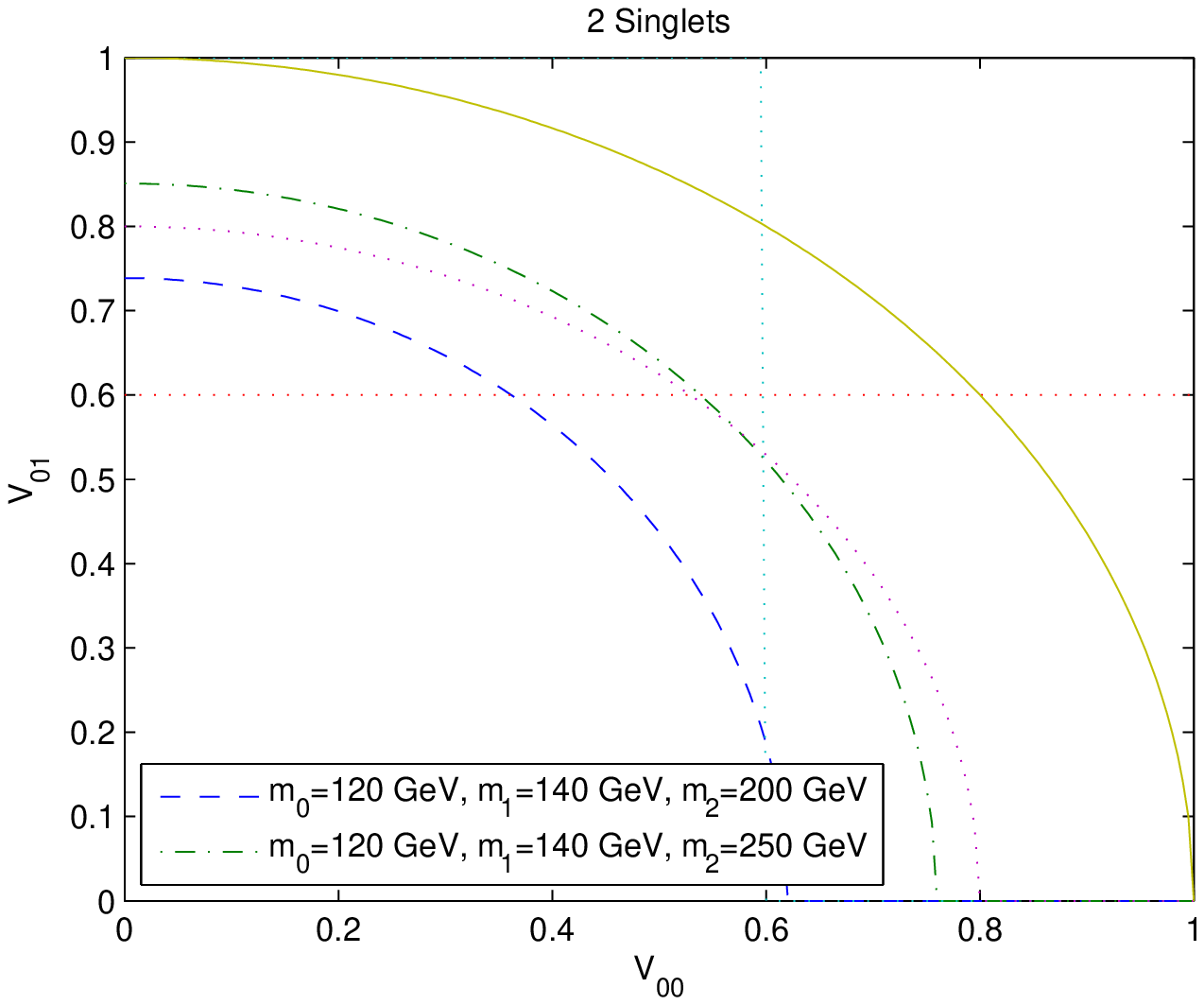}
\caption[]{Allowed region at $95\%$ confidence level for a model
with two singlets in addition to the $SU(2)_L$ scalar doublet.  The
allowed regions are above and to the right of the dashed and
dot-dashed curves. The solid curve is $\sum_i|V_{0i}|^2=1$. 
The solid curve is $\sum_i|V_{0i}|^2=1$. The
curved dotted line is $V_{02}=0.6$, while
the straight dotted lines are $V_{00}=0.6$ and $V_{01}=0.6$.}
\label{sing2b}
\end{center}
\end{figure}

Figures \ref{sing2a} and \ref{sing2b}
show the minimum allowed value
from the electroweak fit for $V_{01}$ as a function of $V_{00}$
for fixed masses. (We assume $\zeta_{ijk}=0$ for simplicity).
The
minimum of $V_{01}$ results from  requiring that the coupling to the
heaviest scalar, $V_{02}=\sqrt{1-V_{00}^2-V_{01}^2}$, not be large
enough that $\phi_2$ makes a significant contribution to
$\Delta S$,$\Delta T$ or $\Delta U$.  The solid red
lines in Figs. \ref{sing2a} and
\ref{sing2b} are  $V_{0i}=0.6$  which roughly represents
the limit of observability in the $\phi_i\rightarrow
ZZ^*\rightarrow 4$~leptons channel.   In these examples,
there is never more than one scalar is observable. In the regions
enclosed by the dotted lines, all three scalars would elude detection
in the $\phi_i\rightarrow Z Z^* \rightarrow 4$~leptons channel with
$30~fb^{-1}$.

The heaviest scalar can have a mass in the $m_2\sim 200-250~GeV$ range
and still have a coupling, $V_{02}$, large enough to be observed in the
$ZZ^*\rightarrow 4$ lepton channel if $m_0$ and $m_1$ are less than
$\sim 160~GeV$, although the lighter scalars will have couplings
which are too small to be observed in this example.  Thus observation
of a scalar Higgs-like particle with $m_i > M_{H,sm}$ can be considered
as a smoking gun for theories with multiple scalar singlets.

\section{Conclusions}
\label{concs}
We have considered the discovery potential for  Higgs bosons
in theories with multiple scalar singlets and demonstrated that quite
simple modifications of the Standard Model Higgs sector can
reduce the significance of the standard Higgs discovery channels.

The addition of two scalar gauge singlets can change the Higgs sector
dramatically from the Standard Model and also from the case with a
single scalar.  In this case it is possible to hide the Higgs boson if
all three physical scalars are light, $m_i \lsim 160~GeV$, with
roughly equal mixing
angles, $V_{0i}\sim {1/\sqrt{3}}$.  Alternatively, in this case, the
electroweak precision measurements allow a Higgs boson in the
$200-250~GeV$ mass
region with couplings to Standard Model particles which are large enough
to allow discovery.

\section*{Acknowledgements}
The work of S.D. is supported by the U.S. Department of Energy under grant
DE-AC02-98CH10886.

\section*{Appendix A}
The Passarino-Veltman functions\cite{Passarino:1978jh}, are defined as,
\begin{eqnarray}
{i\over 16\pi^2}B_0(q^2,m_1,m_2)&=&\int {d^nk\over (2\pi)^n}{1\over
[k^2-m_1^2][(k+q)^2-m_2^2]}\nonumber \\
{i\over 16\pi^2} \biggl\{ g^{\mu\nu}B_{22}(q^2,m_1,m_2) +q^\mu q^\nu
B_{12}(q^2,m_1,m_2)\biggr\} &=&\int {d^nk\over (2\pi)^n}{k^\mu
k^\nu\over [k^2-m_1^2][(k+q)^2-m_2^2]}
\nonumber \\
\end{eqnarray}
We define,
\begin{eqnarray}
B_0(p^2,m_1,m_2)&=&[N_2]\biggl\{{1\over\epsilon}-F_1(p^2,m_1,m_2)\biggr\}
\nonumber \\
B_{22}(p^2,m_1,m_2)&=&[N_2]
m_1^2\biggl\{
{1+r\over 4}\biggl({1\over  \epsilon}+1\biggr)
-{p^2\over 12 m_1^2}\biggl({1\over\epsilon}+1\biggr)
-{1\over 2}F_2(p^2,m_1,m_2)
\biggr\}
\end{eqnarray}
where
\begin{eqnarray}
F_1(p^2,m_1,m_2)&\equiv & \int_0^1dx\ln \biggl( 1-x+{x\over
r}-{p^2x(1-x)\over m_2^2} \biggr)
\nonumber \\
F_{2}(p^2,m_1,m_2)&\equiv& \int_0^1dx \biggl\{ (1-x)+r x-{p^2\over
m_1^2}x(1-x) \biggr\} \ln \biggl( x+{(1-x)\over r} -{p^2\over
m_2^2}x(1-x) \biggr) .
\end{eqnarray}
We need the special cases\cite{Kniehl:1993ay},
\begin{eqnarray}
F_1(0,m_1,m_2)&=&-1-{1\over 1-r}\ln (r)
\nonumber \\
F_1(m_2^2,m_1,m_2)&=&-2
-{1\over 2r}\log(r)+{\beta\over 2 r}
\ln\biggl(
{1-\beta\over 1+\beta}
\biggr)\nonumber \\
F_2(0,m_1,m_2)&=&-{1+r\over 4}-{1\over 2(1-r)}\ln (r)
\nonumber \\
F_{2}(m_2^2,m_1,m_2)
&=& {2\over 3}\biggl(1-{1\over 4 r}\biggr)
\biggl\{ 1+F_1(m_2^2,m_1,m_2)\biggr\} -{1\over 6 r}\log
(r)
-\biggl({1\over 3}+{2 r\over 9}\biggr)\, ,
\end{eqnarray}
and we define
\begin{eqnarray}
r&\equiv& {m_2^2\over m_1^2}
\nonumber \\
\beta&\equiv&\sqrt{1-4r}
\nonumber \\
%\end{eqnarray}
%and
%\begin{equation}
\biggl[ N_2\biggr]&\equiv &
\biggl(
{4\pi\mu^2\over m_2^2}
\biggr)^\epsilon
\Gamma(1+\epsilon) \, .
\end{eqnarray}
\section*{Appendix B}
We use the fit to electroweak
precision data given in Ref. \cite{Profumo:2007wc},
\begin{eqnarray}
\Delta S = S-S_{SM}&=& -0.126 \pm 0.096 \nonumber \\
\Delta T = T-T_{SM}&=&-0.111\pm 0.109\nonumber \\
\Delta U = U-U_{SM}&=&+ 0.164 \pm 0.115
\label{delts}
\end{eqnarray}
with the associated correlation matrix,
\begin{eqnarray}
\rho_{ij}=\left(
\begin{array}{lll}
1.0 & 0.866 & -0.392\nonumber \\
0.866 & 1.0 & -0.588 \nonumber\\
-0.392 & -0.588 & 1.0
 \end{array}
\right)\, .
\end{eqnarray}
$\Delta \chi^2$ is defined as
\begin{equation}
\Delta \chi^2=\Sigma_{ij}(\Delta X_i-\Delta {\hat X}_i)
(\sigma^2)^{-1}_{ij}(\Delta X_i-\Delta {\hat X}_i)\, ,
\end{equation}
where $\Delta {\hat X}_i =\Delta S, \Delta T,$ and
$\Delta U$ are the central values of the fit in Eq.
\ref{delts},  $\Delta X_i=\Delta S_\phi, \Delta T_\phi$, and
$\Delta U_\phi$ from Eq. \ref{stufits},  $\sigma_i$ are the errors
given in Eq. \ref{delts} and $\sigma^2_{ij}=\sigma_i\rho_{ij}\sigma_j$.
The $95\%$ confidence level limit corresponds to $\Delta \chi^2 =
7.815$. 
We vary the input values of $V_{0i}$ and $m_i$ to find the 
$\Delta \chi^2 =
7.815$ contours shown in Figs. 2-5.
 This fit gives a $95\%$ confidence level limit on the Standard
Model Higgs boson of $M_{H,SM} < 166~GeV$.
\bibliography{scalars}

\begin{thebibliography}{32}
\expandafter\ifx\csname natexlab\endcsname\relax\def\natexlab#1{#1}\fi
\expandafter\ifx\csname bibnamefont\endcsname\relax
  \def\bibnamefont#1{#1}\fi
\expandafter\ifx\csname bibfnamefont\endcsname\relax
  \def\bibfnamefont#1{#1}\fi
\expandafter\ifx\csname citenamefont\endcsname\relax
  \def\citenamefont#1{#1}\fi
\expandafter\ifx\csname url\endcsname\relax
  \def\url#1{\texttt{#1}}\fi
\expandafter\ifx\csname urlprefix\endcsname\relax\def\urlprefix{URL }\fi
\providecommand{\bibinfo}[2]{#2}
\providecommand{\eprint}[2][]{\url{#2}}

\bibitem[{\citenamefont{Bahat-Treidel et~al.}(2007)\citenamefont{Bahat-Treidel,
  Grossman, and Rozen}}]{BahatTreidel:2006kx}
\bibinfo{author}{\bibfnamefont{O.}~\bibnamefont{Bahat-Treidel}},
  \bibinfo{author}{\bibfnamefont{Y.}~\bibnamefont{Grossman}}, \bibnamefont{and}
  \bibinfo{author}{\bibfnamefont{Y.}~\bibnamefont{Rozen}},
  \bibinfo{journal}{JHEP} \textbf{\bibinfo{volume}{05}}, \bibinfo{pages}{022}
  (\bibinfo{year}{2007}), \eprint{hep-ph/0611162}.

\bibitem[{\citenamefont{O'Connell et~al.}(2007)\citenamefont{O'Connell,
  Ramsey-Musolf, and Wise}}]{O'Connell:2006wi}
\bibinfo{author}{\bibfnamefont{D.}~\bibnamefont{O'Connell}},
  \bibinfo{author}{\bibfnamefont{M.~J.} \bibnamefont{Ramsey-Musolf}},
  \bibnamefont{and} \bibinfo{author}{\bibfnamefont{M.~B.} \bibnamefont{Wise}},
  \bibinfo{journal}{Phys. Rev.} \textbf{\bibinfo{volume}{D75}},
  \bibinfo{pages}{037701} (\bibinfo{year}{2007}), \eprint{hep-ph/0611014}.

\bibitem[{\citenamefont{Barger et~al.}(2008)\citenamefont{Barger, Langacker,
  McCaskey, Ramsey-Musolf, and Shaughnessy}}]{Barger:2007im}
\bibinfo{author}{\bibfnamefont{V.}~\bibnamefont{Barger}},
  \bibinfo{author}{\bibfnamefont{P.}~\bibnamefont{Langacker}},
  \bibinfo{author}{\bibfnamefont{M.}~\bibnamefont{McCaskey}},
  \bibinfo{author}{\bibfnamefont{M.~J.} \bibnamefont{Ramsey-Musolf}},
  \bibnamefont{and}
  \bibinfo{author}{\bibfnamefont{G.}~\bibnamefont{Shaughnessy}},
  \bibinfo{journal}{Phys. Rev.} \textbf{\bibinfo{volume}{D77}},
  \bibinfo{pages}{035005} (\bibinfo{year}{2008}), \eprint{0706.4311}.

\bibitem[{\citenamefont{Profumo et~al.}(2007)\citenamefont{Profumo,
  Ramsey-Musolf, and Shaughnessy}}]{Profumo:2007wc}
\bibinfo{author}{\bibfnamefont{S.}~\bibnamefont{Profumo}},
  \bibinfo{author}{\bibfnamefont{M.~J.} \bibnamefont{Ramsey-Musolf}},
  \bibnamefont{and}
  \bibinfo{author}{\bibfnamefont{G.}~\bibnamefont{Shaughnessy}},
  \bibinfo{journal}{JHEP} \textbf{\bibinfo{volume}{08}}, \bibinfo{pages}{010}
  (\bibinfo{year}{2007}), \eprint{0705.2425}.

\bibitem[{\citenamefont{Bhattacharyya et~al.}(2008)\citenamefont{Bhattacharyya,
  Branco, and Nandi}}]{Bhattacharyya:2007pb}
\bibinfo{author}{\bibfnamefont{G.}~\bibnamefont{Bhattacharyya}},
  \bibinfo{author}{\bibfnamefont{G.~C.} \bibnamefont{Branco}},
  \bibnamefont{and} \bibinfo{author}{\bibfnamefont{S.}~\bibnamefont{Nandi}},
  \bibinfo{journal}{Phys. Rev.} \textbf{\bibinfo{volume}{D77}},
  \bibinfo{pages}{117701} (\bibinfo{year}{2008}), \eprint{0712.2693}.

\bibitem[{\citenamefont{Barger et~al.}(2009)\citenamefont{Barger, Langacker,
  McCaskey, Ramsey-Musolf, and Shaughnessy}}]{Barger:2008jx}
\bibinfo{author}{\bibfnamefont{V.}~\bibnamefont{Barger}},
  \bibinfo{author}{\bibfnamefont{P.}~\bibnamefont{Langacker}},
  \bibinfo{author}{\bibfnamefont{M.}~\bibnamefont{McCaskey}},
  \bibinfo{author}{\bibfnamefont{M.}~\bibnamefont{Ramsey-Musolf}},
  \bibnamefont{and}
  \bibinfo{author}{\bibfnamefont{G.}~\bibnamefont{Shaughnessy}},
  \bibinfo{journal}{Phys. Rev.} \textbf{\bibinfo{volume}{D79}},
  \bibinfo{pages}{015018} (\bibinfo{year}{2009}), \eprint{0811.0393}.

\bibitem[{\citenamefont{Hewett and Rizzo}(2003)}]{Hewett:2002nk}
\bibinfo{author}{\bibfnamefont{J.~L.} \bibnamefont{Hewett}} \bibnamefont{and}
  \bibinfo{author}{\bibfnamefont{T.~G.} \bibnamefont{Rizzo}},
  \bibinfo{journal}{JHEP} \textbf{\bibinfo{volume}{08}}, \bibinfo{pages}{028}
  (\bibinfo{year}{2003}), \eprint{hep-ph/0202155}.

\bibitem[{\citenamefont{Barger et~al.}(2006)\citenamefont{Barger, Langacker,
  Lee, and Shaughnessy}}]{Barger:2006dh}
\bibinfo{author}{\bibfnamefont{V.}~\bibnamefont{Barger}},
  \bibinfo{author}{\bibfnamefont{P.}~\bibnamefont{Langacker}},
  \bibinfo{author}{\bibfnamefont{H.-S.} \bibnamefont{Lee}}, \bibnamefont{and}
  \bibinfo{author}{\bibfnamefont{G.}~\bibnamefont{Shaughnessy}},
  \bibinfo{journal}{Phys. Rev.} \textbf{\bibinfo{volume}{D73}},
  \bibinfo{pages}{115010} (\bibinfo{year}{2006}), \eprint{hep-ph/0603247}.

\bibitem[{\citenamefont{Dermisek and Gunion}(2006)}]{Dermisek:2005gg}
\bibinfo{author}{\bibfnamefont{R.}~\bibnamefont{Dermisek}} \bibnamefont{and}
  \bibinfo{author}{\bibfnamefont{J.~F.} \bibnamefont{Gunion}},
  \bibinfo{journal}{Phys. Rev.} \textbf{\bibinfo{volume}{D73}},
  \bibinfo{pages}{111701} (\bibinfo{year}{2006}), \eprint{hep-ph/0510322}.

\bibitem[{\citenamefont{Dermisek and Gunion}(2005)}]{Dermisek:2005ar}
\bibinfo{author}{\bibfnamefont{R.}~\bibnamefont{Dermisek}} \bibnamefont{and}
  \bibinfo{author}{\bibfnamefont{J.~F.} \bibnamefont{Gunion}},
  \bibinfo{journal}{Phys. Rev. Lett.} \textbf{\bibinfo{volume}{95}},
  \bibinfo{pages}{041801} (\bibinfo{year}{2005}), \eprint{hep-ph/0502105}.

\bibitem[{\citenamefont{Ellis et~al.}(1989)\citenamefont{Ellis, Gunion, Haber,
  Roszkowski, and Zwirner}}]{Ellis:1988er}
\bibinfo{author}{\bibfnamefont{J.~R.} \bibnamefont{Ellis}},
  \bibinfo{author}{\bibfnamefont{J.~F.} \bibnamefont{Gunion}},
  \bibinfo{author}{\bibfnamefont{H.~E.} \bibnamefont{Haber}},
  \bibinfo{author}{\bibfnamefont{L.}~\bibnamefont{Roszkowski}},
  \bibnamefont{and} \bibinfo{author}{\bibfnamefont{F.}~\bibnamefont{Zwirner}},
  \bibinfo{journal}{Phys. Rev.} \textbf{\bibinfo{volume}{D39}},
  \bibinfo{pages}{844} (\bibinfo{year}{1989}).

\bibitem[{\citenamefont{Bowen et~al.}(2007)\citenamefont{Bowen, Cui, and
  Wells}}]{Bowen:2007ia}
\bibinfo{author}{\bibfnamefont{M.}~\bibnamefont{Bowen}},
  \bibinfo{author}{\bibfnamefont{Y.}~\bibnamefont{Cui}}, \bibnamefont{and}
  \bibinfo{author}{\bibfnamefont{J.~D.} \bibnamefont{Wells}},
  \bibinfo{journal}{JHEP} \textbf{\bibinfo{volume}{03}}, \bibinfo{pages}{036}
  (\bibinfo{year}{2007}), \eprint{hep-ph/0701035}.

\bibitem[{\citenamefont{Schabinger and Wells}(2005)}]{Schabinger:2005ei}
\bibinfo{author}{\bibfnamefont{R.}~\bibnamefont{Schabinger}} \bibnamefont{and}
  \bibinfo{author}{\bibfnamefont{J.~D.} \bibnamefont{Wells}},
  \bibinfo{journal}{Phys. Rev.} \textbf{\bibinfo{volume}{D72}},
  \bibinfo{pages}{093007} (\bibinfo{year}{2005}), \eprint{hep-ph/0509209}.

\bibitem[{\citenamefont{Patt and Wilczek}(2006)}]{Patt:2006fw}
\bibinfo{author}{\bibfnamefont{B.}~\bibnamefont{Patt}} \bibnamefont{and}
  \bibinfo{author}{\bibfnamefont{F.}~\bibnamefont{Wilczek}}
  (\bibinfo{year}{2006}), \eprint{hep-ph/0605188}.

\bibitem[{\citenamefont{Strassler and Zurek}(2007)}]{Strassler:2006im}
\bibinfo{author}{\bibfnamefont{M.~J.} \bibnamefont{Strassler}}
  \bibnamefont{and} \bibinfo{author}{\bibfnamefont{K.~M.} \bibnamefont{Zurek}},
  \bibinfo{journal}{Phys. Lett.} \textbf{\bibinfo{volume}{B651}},
  \bibinfo{pages}{374} (\bibinfo{year}{2007}), \eprint{hep-ph/0604261}.

\bibitem[{\citenamefont{Chang et~al.}(2008)\citenamefont{Chang, Dermisek,
  Gunion, and Weiner}}]{Chang:2008cw}
\bibinfo{author}{\bibfnamefont{S.}~\bibnamefont{Chang}},
  \bibinfo{author}{\bibfnamefont{R.}~\bibnamefont{Dermisek}},
  \bibinfo{author}{\bibfnamefont{J.~F.} \bibnamefont{Gunion}},
  \bibnamefont{and} \bibinfo{author}{\bibfnamefont{N.}~\bibnamefont{Weiner}},
  \bibinfo{journal}{Ann. Rev. Nucl. Part. Sci.} \textbf{\bibinfo{volume}{58}},
  \bibinfo{pages}{75} (\bibinfo{year}{2008}), \eprint{0801.4554}.

\bibitem[{\citenamefont{Barate et~al.}(2003)}]{Barate:2003sz}
\bibinfo{author}{\bibfnamefont{R.}~\bibnamefont{Barate}} \bibnamefont{et~al.}
  (\bibinfo{collaboration}{LEP Working Group for Higgs boson searches}),
  \bibinfo{journal}{Phys. Lett.} \textbf{\bibinfo{volume}{B565}},
  \bibinfo{pages}{61} (\bibinfo{year}{2003}), \eprint{hep-ex/0306033}.

\bibitem[{\citenamefont{Amsler et~al.}(2008)}]{Amsler:2008zzb}
\bibinfo{author}{\bibfnamefont{C.}~\bibnamefont{Amsler}} \bibnamefont{et~al.}
  (\bibinfo{collaboration}{Particle Data Group}), \bibinfo{journal}{Phys.
  Lett.} \textbf{\bibinfo{volume}{B667}}, \bibinfo{pages}{1}
  (\bibinfo{year}{2008}).

\bibitem[{\citenamefont{{LEP Electroweak Working Group}}()}]{lepeww}
\bibinfo{author}{\bibnamefont{{LEP Electroweak Working Group}}},
  \bibinfo{note}{http://lepewwg.web.cern.ch/LEPEWWG/}.

\bibitem[{\citenamefont{Erler and Langacker}(2008)}]{Erler:2008ek}
\bibinfo{author}{\bibfnamefont{J.}~\bibnamefont{Erler}} \bibnamefont{and}
  \bibinfo{author}{\bibfnamefont{P.}~\bibnamefont{Langacker}}
  (\bibinfo{year}{2008}), \eprint{0807.3023}.

\bibitem[{\citenamefont{Zhang et~al.}(2008)\citenamefont{Zhang, Yan, and
  Li}}]{Zhang:2006vt}
\bibinfo{author}{\bibfnamefont{H.-H.} \bibnamefont{Zhang}},
  \bibinfo{author}{\bibfnamefont{W.-B.} \bibnamefont{Yan}}, \bibnamefont{and}
  \bibinfo{author}{\bibfnamefont{X.-S.} \bibnamefont{Li}},
  \bibinfo{journal}{Mod. Phys. Lett.} \textbf{\bibinfo{volume}{A23}},
  \bibinfo{pages}{637} (\bibinfo{year}{2008}), \eprint{hep-ph/0612059}.

\bibitem[{\citenamefont{Dugan and Randall}(1991)}]{Dugan:1991ck}
\bibinfo{author}{\bibfnamefont{M.~J.} \bibnamefont{Dugan}} \bibnamefont{and}
  \bibinfo{author}{\bibfnamefont{L.}~\bibnamefont{Randall}},
  \bibinfo{journal}{Phys. Lett.} \textbf{\bibinfo{volume}{B264}},
  \bibinfo{pages}{154} (\bibinfo{year}{1991}).

\bibitem[{\citenamefont{Chen et~al.}(2006)\citenamefont{Chen, Dawson, and
  Krupovnickas}}]{Chen:2006pb}
\bibinfo{author}{\bibfnamefont{M.-C.} \bibnamefont{Chen}},
  \bibinfo{author}{\bibfnamefont{S.}~\bibnamefont{Dawson}}, \bibnamefont{and}
  \bibinfo{author}{\bibfnamefont{T.}~\bibnamefont{Krupovnickas}},
  \bibinfo{journal}{Phys. Rev.} \textbf{\bibinfo{volume}{D74}},
  \bibinfo{pages}{035001} (\bibinfo{year}{2006}), \eprint{hep-ph/0604102}.

\bibitem[{\citenamefont{Peskin and Takeuchi}(1992)}]{Peskin:1991sw}
\bibinfo{author}{\bibfnamefont{M.~E.} \bibnamefont{Peskin}} \bibnamefont{and}
  \bibinfo{author}{\bibfnamefont{T.}~\bibnamefont{Takeuchi}},
  \bibinfo{journal}{Phys. Rev.} \textbf{\bibinfo{volume}{D46}},
  \bibinfo{pages}{381} (\bibinfo{year}{1992}).

\bibitem[{\citenamefont{Altarelli and Barbieri}(1991)}]{Altarelli:1990zd}
\bibinfo{author}{\bibfnamefont{G.}~\bibnamefont{Altarelli}} \bibnamefont{and}
  \bibinfo{author}{\bibfnamefont{R.}~\bibnamefont{Barbieri}},
  \bibinfo{journal}{Phys. Lett.} \textbf{\bibinfo{volume}{B253}},
  \bibinfo{pages}{161} (\bibinfo{year}{1991}).

\bibitem[{\citenamefont{Degrassi et~al.}(1993)\citenamefont{Degrassi, Kniehl,
  and Sirlin}}]{Degrassi:1993kn}
\bibinfo{author}{\bibfnamefont{G.}~\bibnamefont{Degrassi}},
  \bibinfo{author}{\bibfnamefont{B.~A.} \bibnamefont{Kniehl}},
  \bibnamefont{and} \bibinfo{author}{\bibfnamefont{A.}~\bibnamefont{Sirlin}},
  \bibinfo{journal}{Phys. Rev.} \textbf{\bibinfo{volume}{D48}},
  \bibinfo{pages}{3963} (\bibinfo{year}{1993}).

\bibitem[{\citenamefont{Chen et~al.}(2008)\citenamefont{Chen, Dawson, and
  Jackson}}]{Chen:2008jg}
\bibinfo{author}{\bibfnamefont{M.-C.} \bibnamefont{Chen}},
  \bibinfo{author}{\bibfnamefont{S.}~\bibnamefont{Dawson}}, \bibnamefont{and}
  \bibinfo{author}{\bibfnamefont{C.~B.} \bibnamefont{Jackson}},
  \bibinfo{journal}{Phys. Rev.} \textbf{\bibinfo{volume}{D78}},
  \bibinfo{pages}{093001} (\bibinfo{year}{2008}), \eprint{0809.4185}.

\bibitem[{\citenamefont{Hollik}(1990)}]{Hollik:1988ii}
\bibinfo{author}{\bibfnamefont{W.~F.~L.} \bibnamefont{Hollik}},
  \bibinfo{journal}{Fortschr. Phys.} \textbf{\bibinfo{volume}{38}},
  \bibinfo{pages}{165} (\bibinfo{year}{1990}).

\bibitem[{\citenamefont{Bayatian et~al.}(2006)}]{Bayatian:2006zz}
\bibinfo{author}{\bibfnamefont{G.~L.} \bibnamefont{Bayatian}}
  \bibnamefont{et~al.} (\bibinfo{collaboration}{CMS}) (\bibinfo{year}{2006}),
  \bibinfo{note}{cERN-LHCC-2006-001}.

\bibitem[{\citenamefont{Aad et~al.}(2009)}]{Aad:2009wy}
\bibinfo{author}{\bibfnamefont{G.}~\bibnamefont{Aad}} \bibnamefont{et~al.}
  (\bibinfo{collaboration}{The ATLAS}) (\bibinfo{year}{2009}),
  \eprint{0901.0512}.

\bibitem[{\citenamefont{Passarino and Veltman}(1979)}]{Passarino:1978jh}
\bibinfo{author}{\bibfnamefont{G.}~\bibnamefont{Passarino}} \bibnamefont{and}
  \bibinfo{author}{\bibfnamefont{M.~J.~G.} \bibnamefont{Veltman}},
  \bibinfo{journal}{Nucl. Phys.} \textbf{\bibinfo{volume}{B160}},
  \bibinfo{pages}{151} (\bibinfo{year}{1979}).

\bibitem[{\citenamefont{Kniehl}(1994)}]{Kniehl:1993ay}
\bibinfo{author}{\bibfnamefont{B.~A.} \bibnamefont{Kniehl}},
  \bibinfo{journal}{Phys. Rept.} \textbf{\bibinfo{volume}{240}},
  \bibinfo{pages}{211} (\bibinfo{year}{1994}).

\end{thebibliography}
\end{document}